\documentclass[aip,amsmath,amssymb,floatfix,citeautoscript,reprint]{revtex4-2}
\usepackage{cancel}
\usepackage{amsmath}
\usepackage{graphicx}
\usepackage{amssymb}
\usepackage{amsthm}
\usepackage{bm}
\usepackage{dcolumn}% Align table columns on decimal point
\usepackage{braket}% Enables braket notation
\usepackage{longtable}
\usepackage{siunitx}

\usepackage{ragged2e}
\usepackage{txfonts}

\usepackage{color}
\usepackage[usenames,dvipsnames]{xcolor}
\definecolor{myblue}{rgb}{0,0,1}
\usepackage[breaklinks=true,colorlinks=true,linkcolor=myblue,urlcolor=myblue,citecolor=myblue]{hyperref}
\makeatletter
\renewcommand*{\@textcolor}[3]{%
  \protect\leavevmode
  \begingroup
    \color#1{#2}#3%
  \endgroup
}
\makeatother

\newcommand{\cbcn}{C\textsubscript{B}C\textsubscript{N}~}
\newcommand{\ov}{O\textsubscript{V}~}
\newcommand{\cv}{C\textsubscript{V}~}

\begin{document}

\title{Optical properties of defects in solids via quantum embedding with good active space orbitals}

\author{Bryan T. G. Lau}
\affiliation{Department of Chemistry, Columbia University, New York, New York 10027 USA}
\affiliation{Center for Computational Quantum Physics, Flatiron Institute, New York, New York 10010 USA}
\author{Brian Busemeyer}
\affiliation{Center for Computational Quantum Physics, Flatiron Institute, New York, New York 10010 USA}
\author{Timothy C. Berkelbach}
\email{t.berkelbach@columbia.edu}
\affiliation{Department of Chemistry, Columbia University, New York, New York 10027 USA}
\affiliation{Center for Computational Quantum Physics, Flatiron Institute, New York, New York 10010 USA}

\begin{abstract}

The study of isolated defects in solids is a natural target for classical or
quantum embedding methods that treat the defect at a high level of theory and
the rest of the solid at a lower level of theory. Here, in the context of
active-space-based quantum embeddings, we study the performance of three
active-space orbital selection schemes based on canonical (energy-ordered)
orbitals, local orbitals defined in the spirit of density matrix embedding
theory, and approximate natural transition orbitals.  Using equation-of-motion
coupled-cluster theory with single and double excitations (CCSD), we apply
these active space selection schemes to the calculation of the vertical singlet
excitation energy of a substitutional carbon dimer defect in hexagonal boron
nitride, an oxygen vacancy in magnesium oxide, and a carbon vacancy in diamond.
Especially when used in combination with a simple composite correction,
we find that the best performing schemes can predict the excitation energy
to about 0.1--0.2~eV of its converged value using only a few hundred orbitals,
even when the full supercell has thousands of orbitals, which amounts
to many-orders-of-magnitude computational savings when using correlated
electronic structure theories.
When compared to assigned experimental spectra and accounting for vibrational
corrections, we find that CCSD predicts excitation energies that are accurate
to about 0.1--0.3~eV, which is comparable to its performance in molecules
and bulk solids.
\end{abstract}

\maketitle

\section{Introduction}
\label{sec:intro}

Point defects in solids are a promising platform for single photon emission and
quantum information due to their unique electronic structure and associated
optical properties~\cite{Jelezko2006,Weber2010,Tran2016}.  The computational
study of the electronic structure of defects in solids requires the use of
large clusters to eliminate boundary effects or large supercells to eliminate
spurious periodic
interactions~\cite{Hine2009,Freysoldt2009,Freysoldt2014,Gali2019}.  These
finite clusters or periodic supercells, which may contain hundreds of atoms and
thousands of electrons, can be readily studied with density functional theory
(DFT) but are much more difficult to study via correlated methods, many of
which have computational costs that scale as a high power or exponentially with
the number of electrons and basis functions.  The high computational cost and
local character of point defects motivate the development of various classical
or quantum embedding
methods~\cite{Erbetta2000,Chulhai2018,Eskridge2019,Reimers2020,Gallo2021,Schaefer2021,Lau2021,Ma2021,Mitra2021,Muechler2022},
which seek to approximate the result of the full supercell without an explicit
calculation at the high level of theory.  Analogous observations on large
molecules have motivated similar embedding
strategies~\cite{Kluener2002,Khait2010,List2016,Wen2019,Tran2019,Ye2021}.

The use of an active space is one of the simplest types of quantum embedding,
where a mean-field calculation (Hartree-Fock or DFT) on the full system is
followed by a correlated calculation with a truncated number of electrons and
orbitals.  Active-space-based quantum embedding methods are differentiated by
the choice of active space orbitals, the enforcement of self-consistency
between high-level and low-level calculations, and the method by which
electrons and orbitals outside of the active space are accounted for (if at
all).  For example, one popular approach to the latter uses the constrained
random-phase approximation (cRPA)~\cite{Aryasetiawan2004,Ma2021,Muechler2022}
to generate screened Coulomb interactions that are used in the high-level
active space calculation.

The present work is concerned with the optimal choice of active space orbitals
for neutral excited-state calculations, although much of our discussion applies
to the general electronic structure of defects in solids.  An obvious route to
systematic improvement in quantum embeddings is to grow the size of the active
space.  A ``good'' method to choose active space orbitals is one whose results
converge quickly with increasing active space size.  However, the cost of
common high-level solvers, such as full configuration interaction (also known
as exact diagonalization) precludes a systematic study of this convergence
behavior and, therefore, obscures the optimal choice of active space orbitals.

Here, we calculate defect excitation energies using periodic equation-of-motion
coupled-cluster theory with single and double excitations
(CCSD)~\cite{Stanton1993,Krylov2008} as the high-level active space method.
Despite storage and compute costs that scale as $N^4$ and $N^6$,
respectively---precluding direct application to large supercells in quality
basis sets---CCSD calculations can be routinely performed with about 1000
orbitals, allowing us to go systematically beyond minimal active spaces,
explore the rate of convergence towards full supercell results, and thereby
compare the quality of different methods for choosing active space orbitals.
Recent work from our group has found that CCSD yields neutral excitation
energies of bulk semiconductors and insulators that are accurate to about
0.3~eV~\cite{Wang2020,Wang2021}, motivating the present study of defect
excitation energies.

\section{Methods}
\label{sec:methods}

All of our calculations are performed with a single periodic Hartree-Fock (HF)
reference determinant.  The active space problem is defined using the true
Hamiltonian without any effective one- or two-body operators (aside from the
usual frozen-core potential), i.e., we require no double-counting corrections
and we do not perform any downfolding of transitions outside the active space
as done in other
frameworks~\cite{Kowalski2018,Callahan2021,Ma2021,Muechler2022}; in the context
of embedding, this choice mirrors the philosophy of so-called ``full cell''
dynamical mean-field theory~\cite{Zhu2021}. The occupied and unoccupied orbital
active spaces are defined separately so as not to alter the HF reference.
Specifically, we test the use of three active space schemes based on canonical
orbitals, localized orbitals, and approximate natural transition orbitals,
which are each described next. 

\subsection{Canonical orbitals}
The simple first choice for an active space is a subset of canonical HF
orbitals, i.e., those that diagonalize the Fock matrix.  The most important
orbitals are assumed to be those with energies closest to the band gap. In this
work, the canonical active space is defined by the number of occupied orbitals
(equivalently, the number of electrons being correlated) and the number of
unoccupied orbitals, and convergence is achieved by taking the limit where both
approach the full system size.  Here, for ease of comparison, we use the same
number of occupied and unoccupied orbitals as selected by the local orbital
scheme, described next.  Through testing, we found that this manner of addition
is relatively good, although alternatives can easily be imagined. For example,
information could be incorporated about the character of the canonical
orbitals, prioritizing those with high weight on atoms of interest or with the
necessary symmetry for a given transition.

\subsection{Localized orbitals}

Our second active space choice is motivated by the assumption that defect
physics is local in space.  Although our choice to preserve the HF reference
precludes the use of a popular quantum embedding localizations that mix
occupied and unoccupied orbitals, e.g., with maximally localized Wannier
functions~\cite{Marzari2012,Ma2021,Muechler2022}, here we test a method that is
qualitatively similar.  Specifically, we follow the active space perspective of
density matrix embedding theory~\cite{Knizia2012,Zheng2016} in its regional
embedding formulation, recently introduced by two of us~\cite{Lau2021}.
Regional embedding follows the atomic valence active space
approach~\cite{Sayfutyarova2017}, but targets a larger space of unoccupied
orbitals, which are needed to describe dynamical correlation or electronic
screening.  Regional embedding is also similar to the subsystem projected AO
decomposition approach~\cite{Claudino2019,Claudino2019a}.

We define an operator $\hat{P}_A$ that project onto some atomic orbitals (AOs)
$\mu,\nu$ of a user-selected set of atoms $\{A\}$,
\begin{equation}
\hat{P}_A = \sum_{\mu\nu\in\{A\}} |\mu\rangle [\mathbf{S}]^{-1}_{\mu\nu} \langle \nu|,
\end{equation}
where $\mathbf{S}$ is the overlap matrix.  In regional embedding, the occupied
orbitals are projected onto a minimal set of AOs centered on the atoms in
$\{A\}$, and the virtual orbitals are projected onto the full set of
computational AOs centered on the atoms in $\{A\}$.  The active space is
defined via the eigenvectors of the respective projection operators (which are
rotation matrices of the canonical occupied and unoccupied orbitals) with
eigenvalues above a cutoff.  Definition of the active space thus requires
specification of the atoms to be included in the projection operator and an
eigenvalue cutoff (or equivalently a number of occupied and unoccupied orbitals
to keep). Here we use an eigenvalue cutoff of 0.1 in all results, which
generates unoccupied orbitals spaces that are significantly larger than the
occupied orbital spaces because of their projections onto different sets of
AOs.  In the definition of the projection operator, we include all atoms inside
a sphere centered on the defect, and we grow the active space by increasing the
radius of this sphere.  We will henceforth refer to the active space identified
in this manner as a ``local'' active space.

\subsection{Natural transition orbitals}

Our third and final active space choice is motivated by the successful use of approximate
natural orbitals in ground state calculations. For excited state calculations, the 
analogue is the use of approximate natural transition orbitals (NTOs)~\cite{Martin2003}, 
which have been previously used to reduce the cost of CC calculations of molecular
excited states~\cite{Mata2011,Helmich2011,Baudin2017,Dutta2017,Park2018}.
NTOs are defined via the left
and right singular vectors of the transition density matrix 
$\rho_{pq}^{(n)} = \langle \Psi_0 | a_q^\dagger a_p |\Psi_n\rangle$, where $\Psi_{0/n}$
are the ground and excited states. When the electronic structure does not require
a multireference description, configuration interaction with single excitations (CIS)
provides an economical route to approximate NTOs, which is the approach we use here.
Specifically, the occupied NTO rotation matrix $\mathbf{U}$ and unoccupied NTO rotation matrix $\mathbf{V}$ are obtained
from the singular value decomposition
\begin{equation}
\mathbf{C} = \mathbf{U} \bm{\sigma} \mathbf{V}^\mathrm{T}
\end{equation}
where $|\Psi_n\rangle = \sum_{ai} C_i^a a_a^\dagger a_i |\Phi_0\rangle$ is the CIS
wavefunction of the full supercell with all orbitals. 
The NTO active space is made of pairs of orbitals from $\mathbf{U}$ and $\mathbf{V}$
with the largest singular values $\sigma$. 
Definition of the NTO active space requires a singular value cutoff (or a number of occupied/unoccupied
orbitals to keep).
Because the occupied and unoccupied orbitals are added in pairs, after exhaustion of the
occupied orbital space, we canonicalize the remaining unoccupied orbitals and add them in order
of increasing energy.

\subsection{Correcting for active space size}

As discussed in Sec.~\ref{sec:intro}, quantum embedding results can be improved by approximately
correcting for the finite size of the active space. Instead of downfolding external excitations
into the Hamiltonian, e.g., via the cRPA, we test the performance of 
a simple composite correction. Specifically, we correct the CCSD excitation energy
at the CIS level,
\begin{subequations}
\begin{align}
    E_\mathrm{CCSD}^{\mathrm{full}} &\approx E_\mathrm{CCSD}^{\mathrm{act}}+\delta, \\
    \delta &= E_\mathrm{CIS}^{\mathrm{full}}-E_\mathrm{CIS}^{\mathrm{act}},
\end{align}
\end{subequations}
where ``full'' and ``act'' indicate calculations on the full system and in the
active space.  This correction is affordable due to the low cost of CIS
(comparable to that of the underlying HF calculation), and it is essentially
free when using the NTO active space.  For comparison with our uncorrected CCSD
results, the convergence of the CIS excitation energy in the above three active
spaces is shown in the App.~\ref{app:cis}.  We note that CIS and RPA are
similar theories, and a comparison between composite corrections and
downfolding would be interesting for future work.  Other lower-cost methods can
also be used for the calculation of NTOs or for the composite correction, such
as CC2, partitioned equation-of-motion MP2, or CIS(D)~\cite{Goings2014}.

\section{Results}

We compare the above three active space choices for three representative,
electrically neutral point defects: a \cbcn carbon dimer defect in two-dimensional
hexagonal boron nitride (hBN), an \ov oxygen vacancy in three-dimensional rock salt
MgO (the so-called $F$-center), and a \cv carbon vacancy in three-dimensional carbon
diamond (the so-called GR1 defect). 
Geometry optimizations were performed with Quantum Espresso~\cite{Giannozzi2009} using
density functional theory (DFT), the PBE functional~\cite{Perdew1996}, 
and Kresse-Joubert projector-augmented wave pseudopotentials~\cite{Kresse1999}.
For \cbcn in hBN, we used a kinetic energy cutoff of 70~Ry and a
$4\times 4\times 4$ $k$-point mesh;
for \ov in MgO, we used a kinetic cutoff of 70~Ry and a $9\times 9\times 9$ $k$-point
mesh;
and for \cv in diamond, we used a kinetic energy cutoff of 60~Ry and a
$6\times 6\times 6$ $k$-point mesh.

CIS and CCSD electronic structure calculations were performed at the $\Gamma$
point with PySCF~\cite{Sun2017,Sun2020,McClain2017},
using periodic Gaussian density fitting~\cite{Ye2021a,Ye2021b}.
We employed GTH
pseudopotentials~\cite{Goedecker1996} optimized for HF theory~\cite{Hutter2019}
([He] core for all atoms) and associated correlation-consistent double-zeta
Gaussian basis sets optimized for solids (GTH-cc-pVDZ)~\cite{Ye2022}.  For
vacancy defects, we use a ghost atom with the same basis functions as the
missing atom.  Although we have not exhaustively quantified errors due to the
double-zeta basis, preliminary calculations suggest that these errors are on
the 0.1~eV order of magnitude.  All studied defects have closed-shell,
spin-singlet ground states and, for simplicity, we only study spin-singlet
excitations.  Spin-triplet excitations can be straightforwardly studied using
the same methods, and these are especially interesting in the context of qubit
realization.

\subsection{\cbcn in hBN}

\begin{figure}[b]
\includegraphics[scale=0.9]{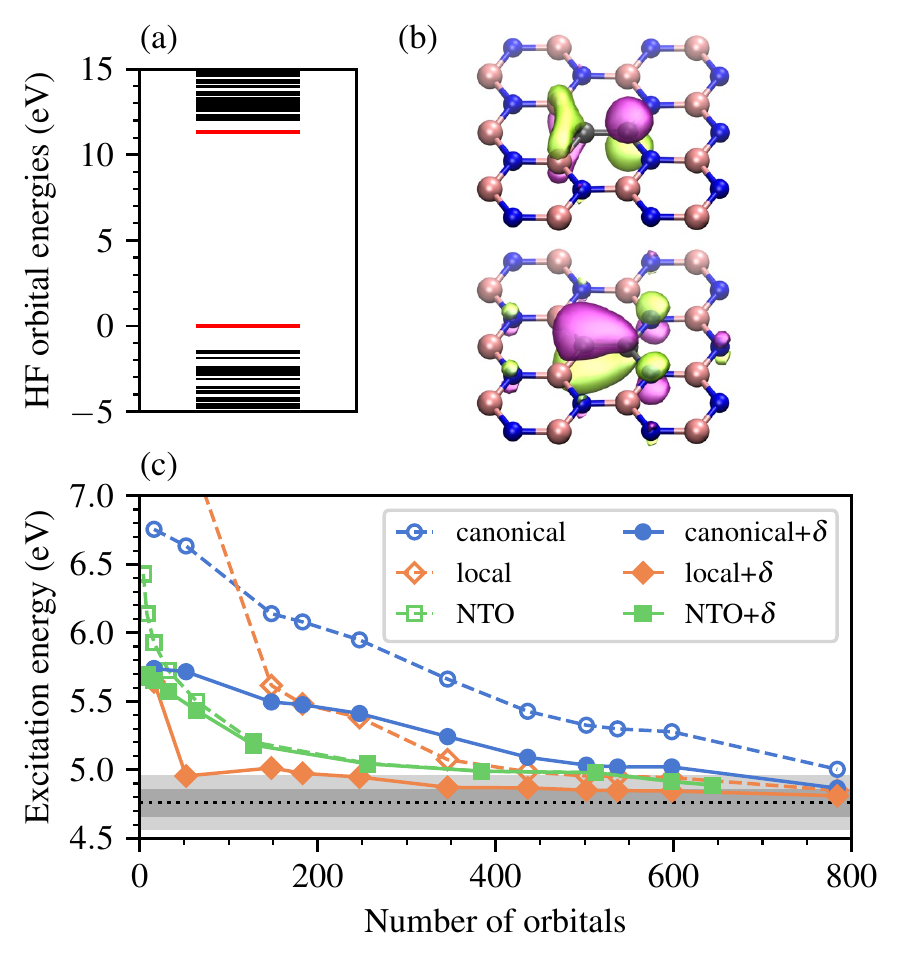}
\caption{
Results for the \cbcn substitutional defect in a $8\times 8\times 1$ supercell of hBN.
(a) The Hartree-Fock molecular orbital energies near the band gap.
Red lines indicate the highest occupied and lowest unoccupied
defect orbitals. (b) The lowest unoccupied (top) and highest occupied (bottom) canonical
HF defect orbitals; only atoms near the defect are shown.
(c) The CCSD excitation energy as a function of active space size for
canonical, local (regional embedding), and CIS NTO active spaces (dashed lines
with open symbols).  Also shown are the results with a composite correction
$\delta$ from a CIS calculation of the full supercell (solid lines with filled
symbols).
Dotted line indicates our best estimate via $M^{-1}$
extrapolation and gray bars are $\pm 0.1$~eV and $\pm 0.2$~eV windows.
}
\label{fig:cchbn}
\end{figure}

Our first example is the \cbcn carbon dimer in hBN.
We study an $8\times 8\times 1$ primitive supercell containing 128 atoms with
20~\textup{\AA} of perpendicular vacuum (which converges the excitation energies).
With our employed basis set and pseudopotentials, the supercell
has 1,664 total orbitals (256 occupied, 1,408 unoccupied), which is slightly
beyond the reach of a routine CCSD calculation.  The canonical valence orbital
energies are shown in Fig.~\ref{fig:cchbn}(a).  The in-gap defect orbitals are
indicated in red and plotted over the atomic structure in
Fig.~\ref{fig:cchbn}(b), and they can be seen to be reasonably localized to the
defect region.  The HF defect orbital gap is 11.32~eV and our best estimate
of the CCSD excitation energy (obtained as described below) is 4.76~eV.

In Fig.~\ref{fig:cchbn}(b), we show the lowest-lying CCSD excitation energy as a
function of the total number of orbitals (occupied plus unoccupied) in the three
active spaces defined in Sec.~\ref{sec:methods}. 
The excitation energy converges fastest in the NTO active space, followed by the
local (regional embedding) active space, followed by the canonical active space.
In the NTO active space, the excitation energy is
converged to 0.2~eV using about 500 orbitals.
The good performance of the NTO active space indicates that the CIS
wavefunction provides a semiquantitative description of the transition density
matrix.
For small active spaces, the local active space yields excitations that are much too high.
However, it performs well for active spaces with more than about 400 orbitals,
giving a result in good agreement with the NTO active space.
The canonical orbital active space is very poor and the results converge
extremely slowly, which is a well-known behavior of wavefunction-based
dynamical correlation.

The CIS composite correction improves the convergence of all methods, but in
qualitatively different ways. Because the NTO basis
is almost perfect for CIS (by construction), the CIS composite correction 
in a truncated NTO active space is very small and does not significantly
impact the CCSD results. In contrast, the CIS composite correction to the
local active space is significant and provides good performance: the corrected 
excitation energy is accurate to 0.2~eV using only 52 orbitals!
The composite correction to results in the canonical active space reduces the
magnitude of the error but does not significantly change the rate of convergence.

\begin{table}[t]
    \centering
    \caption{First singlet excitation energy of the \cbcn defect in hexagonal
boron nitride.
}
    \begin{ruledtabular}
    \label{tab:cchbn}
    \begin{tabular}{lcc}
        Method & Structure & Exc.~energy (eV) \\ 
        \hline
        CCSD (this work) & periodic & 4.76 \\
        TDDFT@CAM-B3LYP~\cite{Korona2019} & cluster & 4.78 \\ % cluster
        $\Delta$SCF@HSE($\alpha=0.4$)~\cite{MackoitSinkeviciene2019} & periodic & 4.53 \\ % pbc
        TDDFT@PBE0~\cite{Winter2021} & cluster & 4.61 \\ % cluster
        evGW+BSE@PBE0~\cite{Winter2021} & cluster & 4.64 \\ %cluster
        cRPA+ED(2e,2o)@PBE~\cite{Muechler2022} & periodic & 3.98 \\ % pbc
        cRPA+ED(2e,2o)@HSE($\alpha=0.4$)~\cite{Muechler2022} & periodic & 4.23 \\ % pbc
        \hline
        Experiment (ZPL)~\cite{Museur2008,Du2015} & --- & 4.1 \\
        Experiment (est.~vertical)~\cite{Winter2021} & --- & 4.6 \\
    \end{tabular}
    \end{ruledtabular}
\end{table}

To estimate the full supercell limit, we extrapolate our composite-corrected result
in the local active space assuming $M^{-1}$ convergence, where $M$ is the number of basis functions.
This yields a vertical excitation energy of 4.76~eV; analogous extrapolation of results
in alternative active spaces gives similar results, differing by less than 0.1~eV.
In Tab.~\ref{tab:cchbn}, we compare this value to previously obtained vertical excitation
energies using 
TDDFT with the CAMB3LYP~\cite{Korona2019} and PBE0~\cite{Winter2021} functionals,
$\Delta$SCF with the HSE functional~\cite{MackoitSinkeviciene2019},
eigenvalue self-consistent GW approximation and Bethe-Salpeter equation (BSE) started from
PBE0~\cite{Winter2021}, and
constrained random-phase approximation (cRPA) screening combined with exact diagonalization (ED)
in a minimal active space of two electrons in two orbitals (2e,2o)~\cite{Muechler2022};
these cRPA+ED results were obtained with both PBE and HSE and include double-counting
corrections that mitigate the dependence on this starting point. In both of the
$\Delta$SCF and cRPA+ED calculations, HSE calculations were performed with the
modified exact-exchange fraction $\alpha=0.4$.
The table indicates whether a finite cluster or periodic structure were used to model
the crystal.

We see that our CCSD prediction is within 0.1--0.2~eV of $\Delta$SCF, TDDFT,
and GW+BSE results, and about 0.6--0.8~eV higher in energy than cRPA+ED
results.  We propose that this system is not especially correlated and does not
require the multiconfigurational treatment of ED; instead, dynamical
correlation is important---as evidenced by the large number of orbitals
required and slow convergence---and cRPA only captures this dynamical
correlation approximately.

Bulk hBN exhibits a zero-phonon line (ZPL) around
4.1~eV~\cite{Museur2008,Du2015} that has been attributed to the optical
absorption of the \cbcn dimer defect~\cite{MackoitSinkeviciene2019},
although we emphasize that this assignment is not definite.  
For direct comparison, lattice reorganization energies, zero-point vibrational
energies, and interlayer or substrate screening effects need to be accounted
for. Previous work has estimated the first two corrections lower the vertical
excitation energy by about
0.2--0.3~eV~\cite{MackoitSinkeviciene2019,Winter2021} and the screening
correction lowers it by another 0.3~eV~\cite{Winter2021}. In this light, our
prediction of 4.76~eV might be an overestimation of only about 0.2~eV, which is
typical for CCSD. 

\subsection{\ov in MgO}

\begin{figure}[b]
\includegraphics[scale=0.9]{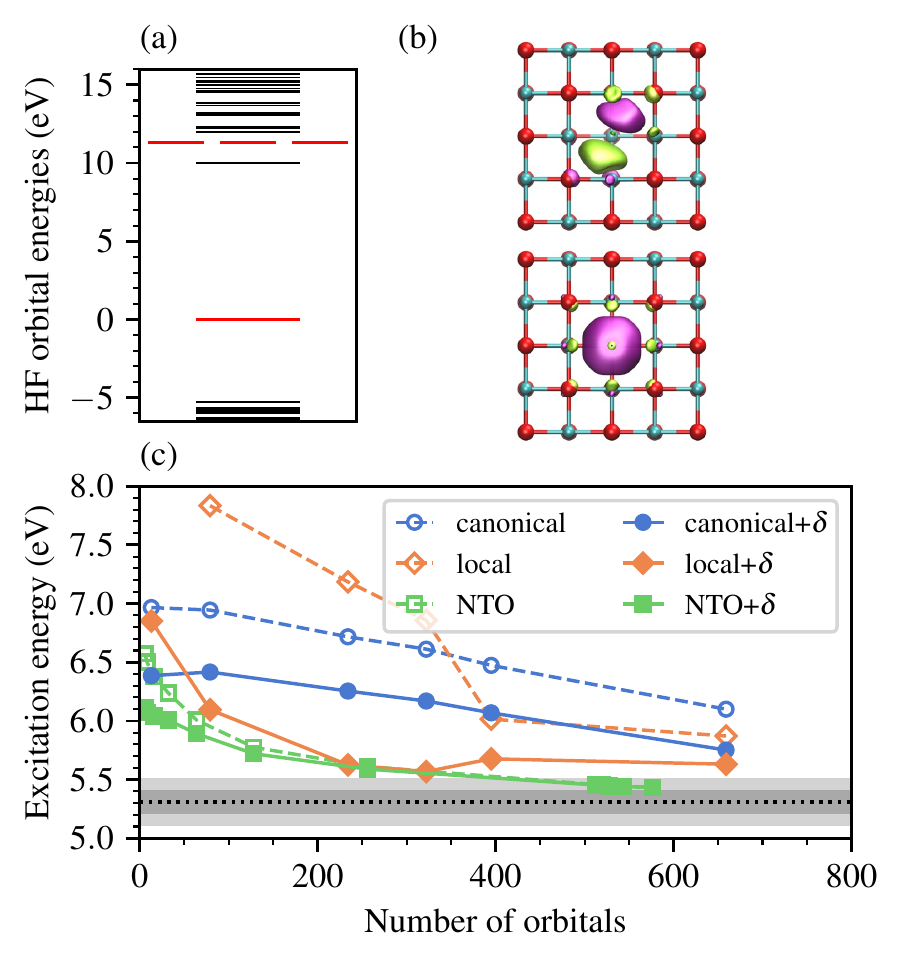}
\caption{
The same as Fig.~\ref{fig:cchbn}, but for the \ov oxygen vacancy ($F$-center) in a
$3\times 3\times 3$ supercell of MgO containing 215 atoms.
Only one of the three degenerate lowest unoccupied defect orbitals is shown.
}
\label{fig:ovmgo}
\end{figure}

Our next example is the \ov oxygen vacancy in MgO, also known as a neutral
$F$-center.  For reference, previous work from our group used periodic CCSD to
estimate the excitation energy of non-defective MgO and found
8.3~eV~\cite{Wang2020}, which can be compared to the experimental value of
7.6~\cite{Roessler1967}.  Electron-phonon calculations predict a zero-point
renormalization of the band gap by
0.3--0.6~eV~\cite{Rinke2012,Antonius2015,Miglio2020}, suggesting that CCSD
overestimates the vertical electronic excitation energy by 0.1--0.4~eV.

We study a $3\times 3\times 3$ conventional supercell containing 215 atoms and
one oxygen ghost atom. 
With our basis set and pseudopotentials, the supercell has 2,916 orbitals
(861 occupied and 2,055 unoccupied), which is much too big for a conventional
CCSD calculation.  The HF molecular orbital energies and defect orbitals are
plotted in Figs.~\ref{fig:ovmgo}(a) and (b), where now the lowest-lying
unoccupied defect orbitals (threefold degenerate) are not in the band gap of
the crystal.  Moreover, this defect orbital is three-fold degenerate due to its
$p$-type symmetry, and the excitations energies are analogously degenerate.
The HF defect gap is 11.33~eV and our best estimate of the CCSD excitation energy
is 5.31~eV.

In Fig.~\ref{fig:ovmgo}(b), we plot the CCSD excitation energy in the same three active space
types, without and with CIS composite corrections.
Unlike for \cbcn, the dimensionality of the MgO crystal prevents us from
achieving converged agreement between various active space choices.  Results in
the NTO active space converge the fastest, and composite corrections are small,
but complete convergence is not observed. Extrapolation of the
composite-corrected NTO results assuming $M^{-1}$ convergence yields our best
prediction of 5.31~eV (again, analogous extrapolation of canonical orbital
results gives good agreement). The uncorrected local orbital results show
a plateau at the two largest active spaces. The same feature is also seen in the
CIS calculation (see App.~\ref{app:cis}), which requires over 1000 orbitals
for convergence.  Although the qualitative similarity explains the relatively
good performance of the composite correction, the cancellation is imperfect and
so convergence is not observed in the corrected local-orbital CCSD
calculations.

\begin{table}[t]
    \centering
    \caption{First singlet excitation energy of the \ov ($F$-center) defect in MgO. 
}
    \begin{ruledtabular}
    \label{tab:ovmgo}
    \begin{tabular}{lcc}
        Method & Structure & Exc.~energy (eV) \\ 
        \hline
        CCSD (this work) & periodic & 5.31 \\
        CASPT2 (2e,2o)~\cite{Sousa2001} & cluster & 5.44 \\
        G$_0$W$_0$+BSE@LDA0~\cite{Rinke2012} & periodic & 4.95 \\
        DMC~\cite{Ertekin2013} & periodic & 5.0 $\pm$ 0.1 \\
        CCSD~\cite{Gallo2021} & periodic & 5.28 \\
        \hline
        Experiment (est.~vertical)~\cite{Chen1969,Kappers1970,Evans1970} & --- & 5.0 \\
    \end{tabular}
    \end{ruledtabular}
\end{table}

In Tab.~\ref{tab:ovmgo}, we compare our singlet excitation energy to that
calculated by complete active space self-consistent field plus second-order
perturbation theory (CASPT2)~\cite{Sousa2001}, G$_0$W$_0$+BSE~\cite{Rinke2012},
diffusion Monte Carlo (DMC)~\cite{Ertekin2013},
and CCSD~\cite{Gallo2021}.  The CASPT2 calculations were done with a 26-atom
cluster model containing the oxygen vacancy and three shells (14 Mg atoms and
12 O atoms), surrounded by ab initio model potentials and an array of optimized
point charges; a minimal active space of two electrons in two orbitals (2e,2o)
was used, and increases to the active space size increased the excitation
energy by about 0.05~eV.  The G$_0$W$_0$+BSE calculations were performed on top
of a LDA reference with a one-shot correction to include 25\% exact exchange
(termed LDA0 in Ref.~\onlinecite{Rinke2012}). The DMC calculations used the fixed-node
approximation with a DFT-based Slater-Jastrow trial wavefunction and supercells containing
up to 64 atoms. Finally, the CCSD
calculations~\cite{Gallo2021}, which are most directly comparable to our own,
were performed for several supercell sizes and basis set truncations and then
extrapolated. For comparison, the largest supercell studied contained 127 atoms
(compared to our 215 atoms) and a canonical orbital active space was used
containing up to about 701 orbitals (comparable to our own), although orbital character
was considered in the construction of the active space.
Our CCSD results and the extrapolated CCSD results from
Ref.~\onlinecite{Gallo2021} agree almost perfectly (0.03~eV difference), 
and both are within about 0.15~eV of the 
CASPT2 results~\cite{Sousa2001}. 

This agreement between different calculations and different methods is
encouraging. Experimental optical measurements exhibit a broad absorption band
with a maximum around 5.0~eV, which has been assigned to the neutral
$F$-center~\cite{Chen1969,Kappers1970}.  Vibronic modeling~\cite{Evans1970}
suggests a ZPL at about 4.0~eV and a reorganization energy of
about 1.0~eV (i.e., a very large Huang-Rhys factor of about 30, which is
consistent with the broad absorption band); therefore, the peak location of
5.0~eV is a reasonable estimation of the true vertical excitation energy.  The
remaining 0.3~eV discrepancy between CCSD and experimental measurements is 
typical of CCSD and consistent with its accuracy for non-defective MgO.

\subsection{\cv in diamond}

\begin{figure}[b]
\includegraphics[scale=0.9]{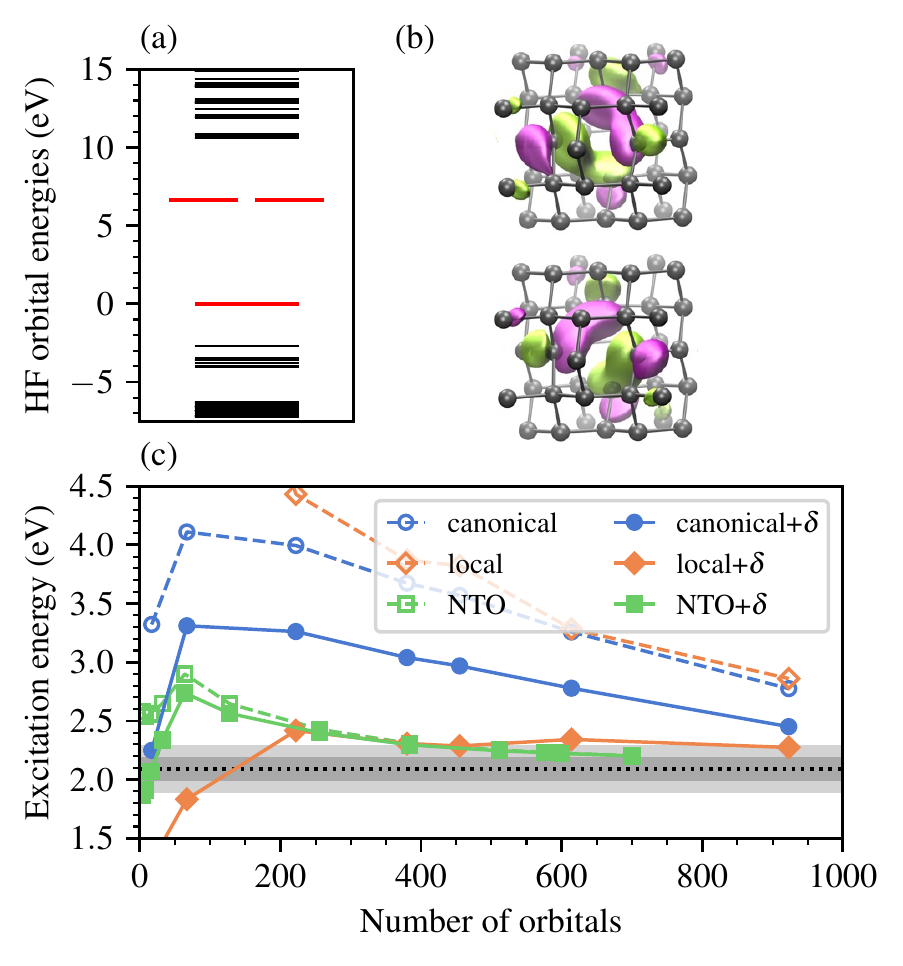}
\caption{
The same as Fig.~\ref{fig:cchbn}, but for the \cv carbon vacancy (GR1 defect) in a
$2\times 2\times 2$ BCC supercell of diamond containing 255 atoms.
Only one of the two degenerate lowest unoccupied defect orbitals is shown.
}
\label{fig:cvc}
\end{figure}

Our final example is the \cv carbon vacancy in diamond.  This particular system
is a valuable prototype of diamond defects, such as the popular negatively
charged nitrogen vacancy center~\cite{Dobrovitski2013}.  Again for reference,
periodic CCSD predicts an excitation of 7.47~eV for non-defective diamond,
which can be compared to the experimental value of 7.3~eV~\cite{Phillip1964}.
Electron-phonon calculations predict a zero-point renormalization of the band
gap by 0.3~eV~\cite{Miglio2020}, suggesting that CCSD underestimates the
excitation energy by about 0.1~eV.

For the \cv vacancy, we study a $2\times 2\times 2$ body-centered cubic
supercell containing 255 carbon atoms and one carbon ghost atom.  With our
employed basis set and pseudopotentials, the system has 3,328 orbitals
(510 occupied and 2,818 unoccupied orbitals), which is our largest system
studied and impossible to treat with conventional CCSD.  
The HF defect gap is 6.63~eV and our best estimate of the CCSD excitation
energy is 2.09~eV.
Before presenting our
results, we first address the nuclear geometry and associated electronic
structure.

Removing a carbon atom from diamond yields a structure with $T_d$ symmetry and
an open shell due to the placement of two valence electrons into a threefold
degenerate set of orbitals with $t_2$ symmetry. This valence suggests a
multiplet structure of low-lying many-body states with multi-configurational
character, which has been studied by fixed-node DMC with a symmetry-adapted
multi-Slater-Jastrow trial wavefunction~\cite{Hood2003}.  Alternatively,
coupling to the lattice can induce a Jahn-Teller
distortion~\cite{Clark1973,Davies1981} that lowers the symmetry to $D_{2d}$,
breaks the threefold degenerate $t_2$ orbitals into one lower $a$ orbital and
two higher $e$ orbitals, and thus yields a closed-shell ground-state singlet
configuration.  While this latter scenario is at odds with experimental support
for tetrahedral symmetry at the vacancy~\cite{Clark1973}, a quantum
\textit{dynamic} Jahn-Teller mechanism is largely supported and has been
studied by anharmonic vibrational calculations~\cite{Prentice2017}.

Because CC theory is only appropriate for single-reference ground states,
we focus on the statically distorted $D_{2d}$ structure, which constitutes one
of the points on the Born-Oppenheimer ground-state surface that is needed to define
the anharmonic vibrational problem.
At the HF level, the orbital energies and defect orbitals are shown in
Figs.~\ref{fig:cvc}(a) and (b). The orbital energies are inside the band gap of
diamond and the lowest-lying unoccupied defect orbital is two-fold degenerate.

The convergence of the CCSD excitation energy is shown in
Fig.~\ref{fig:cvc}(c).  Results obtained with the canonical active space
converge smoothly but slowly, and, as before, composite corrections decrease
the magnitude of the error but do not increase the rate of convergence.
Uncorrected results in the local orbital active space converge the slowest, but
the composite correction is excellent and yields good predictions using only
200--400 orbitals. The NTO results are arguably the best, both without and with
composite corrections. Unlike for \ov in MgO, the excitation energy of \cv in
diamond converges sufficiently quickly that we can achieve agreement between
different active space choices using as few as about 380 orbitals or 10\% of
the total number of orbitals!  Using $M^{-1}$ extrapolation of the corrected
NTO results, our best estimate of the CCSD excitation energy is 2.09~eV.

\begin{table}[t]
    \centering
    \caption{First singlet excitation energies of the \cv (GR1) defect in diamond.
    }
    \begin{ruledtabular}
    \label{tab:cvc}
    \begin{tabular}{lcc}
        Method & Structure & Exc.~energy (eV) \\ 
        \hline
        CCSD (this work) & periodic / $D_{2d}$ & 2.09 \\
        DMC~\cite{Hood2003} & periodic / $T_d$ & 1.51 $\pm$ 0.34 \\
        $\Delta$SCF@B3LYP~\cite{Mackrodt2022} & periodic / pseudo-$T_{d}$ & 1.57 \\
        \hline
        Experiment (ZPL)~\cite{Clark1956} & --- & 1.67 \\
        Experiment (est.~vertical)~\cite{Lannoo1968} & --- & 2.2 \\
    \end{tabular}
    \end{ruledtabular}
\end{table}

In Tab.~\ref{tab:cvc}, we compare our prediction to the multiconfigurational
DMC results in $T_d$ geometry and to very recent $\Delta$SCF@B3LYP results in
a pseudo-$T_d$ geometry (only slight symmetry breaking); the latter breaks spin symmetry
in order to stabilize a single configuration ground state.
The CCSD excitation energy is 0.5--0.6~eV
higher than that of these other calculations.
Experimental low-temperature spectra show a zero-phonon line at
1.67~eV~\cite{Clark1956}, which has been assigned to the \cv (GR1) vacancy.
Vibronic modeling of the statically distorted Jahn-Teller structure predicts a
lattice reorganization energy of about 0.5~eV~\cite{Lannoo1968}, suggesting a
true vertical excitation energy of about 2.2~eV.  The CCSD prediction is
therefore in excellent agreement and, interestingly, an underestimation,
which is the same as in non-defective diamond.

\section{Conclusion}

To summarize, we have explored the slow convergence of excitation energies of
defects in solids with respect to the number of orbitals using CCSD
as a high-level method. In the context
of active-space based quantum embedding, we studied two alternatives to canonical
orbitals---local orbitals and natural transition orbitals---which expedite
the convergence, especially when using composite corrections.
Although achieving precision to better than 0.1~eV still requires
extrapolation, the uncertainty introduced by extrapolation is significantly
reduced. 
For the three defects studied, we find that CCSD predicts vertical 
excitation energies that are accurate to 0.1--0.3~eV when compared to
putative experimental values as long as vibrational corrections are
accounted for. 

We expect the methods introduced here to be applicable to other single-reference
electronic structure methods, such as GW-BSE. 
Future work could aim to enforce self-consistency between the low-level and high-level
treatments of the embedded system, as an improvement to the one-shot embedding
used here.  Extension to defects with multiconfigurational ground states
requires further consideration, both with respect to the active space selection
and the challenge of growing the active space. A similar study of existing
methods, such as cRPA+ED, would be an interesting first step, including the
performance of alternative active spaces and the quantitative treatment of
dynamical correlation by the cRPA.

\section*{Acknowledgments}
We thank Cyrus Dreyer, Xiao Wang, Hong-Zhou Ye, and Shiwei Zhang for helpful
discussions.  This work was supported in part by the US Air Force Office of
Scientific Research under Grant No.~FA9550-21-1-0400.  The Flatiron Institute
is a division of the Simons Foundation.

\appendix
\section{CIS excitation energy convergence}
\label{app:cis}
\begin{figure}[t]
\includegraphics[scale=0.9]{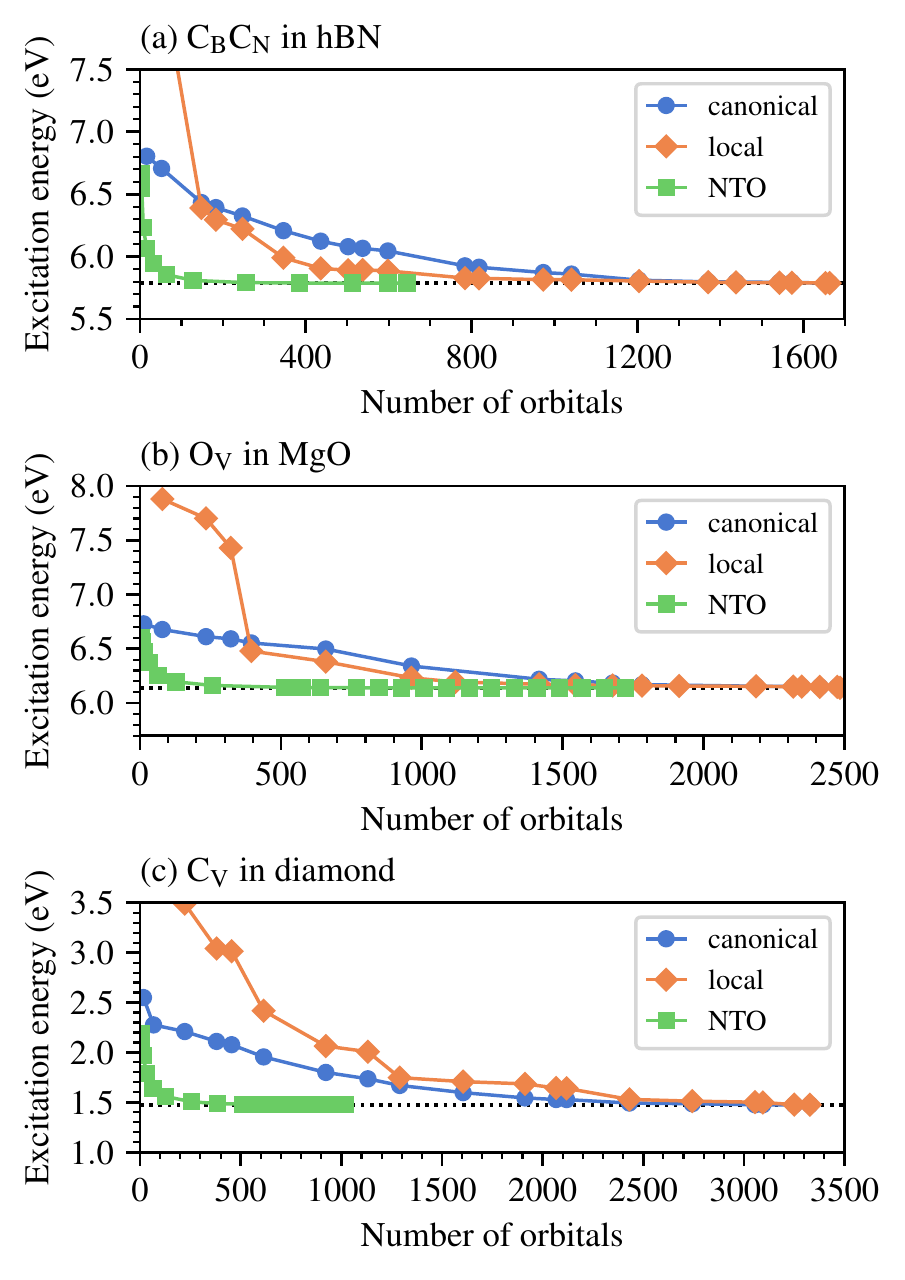}
\caption{CIS excitation energy of the three defects studied in the
main text (indicated in each panel) using the three active space
choices (indicated in the legend).
The final data point of the canonical and local results corresponds
to the full supercell limit in the employed basis.
For visual clarity, NTO results are not shown for large number of orbitals.
}
\label{fig:cis}
\end{figure}

In Fig.~\ref{fig:cis}, we present the CIS excitation energy of the three
defects studied in the main text.

\end{document}